# Simultaneous individual and dipolar collective properties in binary assemblies of magnetic nanoparticles

*Elena H. Sánchez[1*], Marianna Vasilakaki[2], Su Seong Lee[3], Peter S. Normile[1], Giuseppe Muscas[4], Massimiliano Murgia[1,5], Mikael S. Andersson[6,7], Gurvinder Singh[8], Roland Mathieu[6], Per Nordblad[6], Pier Carlo Ricci[5], Davide Peddis[9,10], Kalliopi N. Trohidou[2], Josep Nogués[11,12*], José A. De Toro[1*]*

[1] Instituto Regional de Investigación Científica Aplicada (IRICA) and Departamento de Física Aplicada, Universidad de Castilla-La Mancha 13071 Ciudad Real, Spain

[2] Institute of Nanoscience and Nanotechnology, NCSR "Demokritos", 153 10 Agia Paraskevi, Attiki, Greece

[3] Institute of Bioengineering and Nanotechnology, 31 Biopolis Way, The Nanos, Singapore 138669, Singapore

[4] Department of Physics and Astronomy, Uppsala University, Box 516, SE-751 20, Uppsala, Sweden

[5] Dipartimento di Fisica, Universitá degli Studi di Cagliari, S.P. Monserrato-Sestu Km 0,700, 09042 Monserrato (CA), Italy

[6] Department of Engineering Sciences, Uppsala University, Box 534, SE-751 21 Uppsala, Sweden

[7] Department of Chemistry and Chemical Engineering, Chalmers University of Technology, SE-412 96 Göteborg, Sweden

[8] School of Aerospace, Mechanical and Mechatronic Engineering, University of Sydney, Sydney, NSW 2008, Australia

[9] Università degli Studi di Genova, Dipartimento di Chimica e Chimica Industriale, Via Dodecaneso 31, 1-16146 Genova, Italy

[10] Istituto di Structura della Materia-CNR, 00015 Monterotondo Scalo (RM), Italy

[11] Catalan Institute of Nanoscience and Nanotechnology (ICN2), CSIC and BIST, Campus UAB, Bellaterra, 08193 Barcelona, Spain

[12] ICREA, Pg. Lluís Companys 23, 08010 Barcelona, Spain

**ABSTRACT**

Applications based on aggregates of magnetic nanoparticles are becoming increasingly widespread, ranging from hyperthermia to magnetic recording. However, although some uses require a collective behavior, other need a more individual-like response, the conditions leading to either of these behaviors are still poorly understood. Here we use nanoscale-uniform binary random dense mixtures with different proportions of oxide magnetic nanoparticles with low/high anisotropy as a valuable tool to explore the crossover from individual to collective behavior. Two different *anisotropy scenarios* have been studied in two series of binary compacts: M1, comprising maghemite ($\gamma$-Fe$_2$O$_3$) nanoparticles of different sizes (9.0 nm / 11.5 nm) with barely a factor of 2 between their anisotropy energies and M2, mixing equally-sized pure maghemite (low-anisotropy) and Co-doped maghemite (high-anisotropy) nanoparticles with a large difference in anisotropy energy (ratio > 8). Interestingly, while the M1 series exhibits collective behavior typical of strongly-coupled dipolar systems, the M2 series presents a more complex scenario where different magnetic properties resemble either "individual-like" or "collective", crucially emphasizing that the *collective* character must be ascribed to specific properties and not to the system as a whole. The strong differences between the two series, offer new insight (systematically ratified by simulations) into the subtle interplay between dipolar interactions, local anisotropy and sample heterogeneity, to determine the behavior of dense assemblies of magnetic nanoparticles.

# INTRODUCTION

Dense nanoparticle (NP) assemblies are the basis of an ever-increasing catalogue of applications.[1–4] The advances in synthetic chemistry have allowed the preparation of monodisperse, highly uniform NPs, which in turn has enabled their assembly to build NP analogues of atomic crystals (sometimes called super- or supra-crystals/lattices), either comprising a single type of NP,[5–10] or several species to form supra-compounds exhibiting a remarkable variety of crystal symmetries.[11–14] However, the most studied NP composite systems are disordered mixtures pursuing a combination of properties to optimize a given figure of merit. For instance, in the broad field of nanomagnetism the idea is epitomized by the exchange-coupling strategy between magnetically soft and hard nanograins (with high saturation magnetization and large coercivity, respectively) in order to maximize the energy product of novel permanent magnets.[15–19] These composites are typically metallic and the ferromagnetic grains interact *via* direct exchange, leading to single-phase behavior with enhanced properties.[18–20] On the other hand, compacts of oxide nanoparticles, where the interparticle interactions are mainly of dipole-dipole type, typically show superspin glass behavior[21–23] (previously described for dipolarly-interacting dense ferrofluids[24]).

Collective behavior in magnetic nanoparticle systems can be useful for some applications like hyperthermia and magnetic resonance imaging.[25,26] In the recently-discovered "liquid permanent magnets" based on nanoparticles, strong dipolar interactions are crucial to enhance the thermal stability of the magnetization and transform the droplet surface into a ferromagnetic layer.[27] On the other hand, collective behavior, or even short-range correlations, is detrimental for the performance of magnetic nanoparticles/grains in magnetic storage and magnetoresistance sensing.[28,29] Thus, understanding the collective vs individual behavior of dense systems of nanoparticles becomes crucial to optimize those applications. Although *collective* behavior is a fundamental term in condensed matter physics[30] or any other type of complex network,[31] its meaning is not clear-cut in the context of magnetic nanoparticle systems, where the differentiation between modified-single-particle behavior and collective order driven by dipolar interactions has produced a large body of experimental and theoretical literature.[32–38] In general, the term *collective* is intended to describe the emergence of patterns of large-scale behavior from the complex interactions between small constituent parts. However, different properties are determined at different length scales, prompting the possibility for a given system to exhibit both individual and collective properties. In this context, dense binary assemblies are presented here as a unique tool to shed light on the above ideas, as in these systems "individual" properties will be evidenced by a doublet of values, each corresponding to one type of constituent.

In binary NP composites nanoscale homogeneity is crucial to enable the tuning of the properties beyond that of the simple superposition of the two constituents.[13,14,39–41] Notably, although binary random compacts of oxide NPs, where the local anisotropy can be readily tuned by the proportion of high/low anisotropy particles in the mix, offer the possibility to explore the complex relation between interparticle interactions and local anisotropy, this tool has been very rarely employed to address the issue.[39] Here, exploiting the narrow size distribution of the constituent particles (2% polydispersity), we have prepared what could be considered the simplest possible NP composites by randomly mixing and compacting two populations of NPs with different anisotropy energy barriers. Two complementary series of such mixtures have been prepared with a moderate/large difference in the anisotropy energy of the constituent NPs. The results show that the proportion of low and high anisotropy (LA, HA) particles in these highly homogeneous compacts may be used to fine tune both the hysteresis loops and the low-field magnetization dynamics (e.g., the blocking/freezing temperature) of the assemblies. Moreover, our experiments allow an assessment of the weight of single-particle *versus* dipolar interaction energies in the determination of the above magnetic properties, with contrasting results between the two series studied.

**RESULTS AND DISCUSSION**

The M1 and M2 series of binary compacts were prepared mixing in different proportions highly uniform, roughly spherical, maghemite-based NPs with different size and different anisotropy constant, respectively (see Methods). The particles were mixed while still in liquid solution, which was subsequently dried up, and -after washing out the oleic acid surfactant- the resulting powder was compacted to form dense discs (see inset in Figure 1c). For the M1 series, NPs with mean diameters $d_{TEM}$ = 9.0 and 11.5 nm (corresponding to a volume ratio of 2) were used (Figure 1a,b,d). For M2, equally-sized, 6.8 nm, pure and Co-doped maghemite NPs were mixed (Figure 1e,g). The samples are denoted as M$i$-$x$, where $i$ = 1, 2 refers to the series and $x$ = 0, 10, 20, 30, 50, 65, 85, 100% in both series ($x$, defined as the proportion of HA particles). The uniform mixing of the NPs, down to the particle level, was verified by high resolution scanning electron microscope (HRSEM) images in the case of the M1 series (Figure 1c),[42] and by compositional mapping for the M2 series, where the mixed nanoparticles have the same size but different composition (Figure 1g,h).

In addition, a fraction of the four types of NPs was extracted from each batch to be coated with a thick silica shell ($t \approx 3d_{TEM}$; see *e.g.*, inset in Figure 1b) in order to magnetically isolate the cores and thus measure the single-particle magnetic properties.[22,43] Each type of nanoparticle is characterized by an energy barrier $K_{ef}V$, where $K_{ef}$ is the effective anisotropy and V is the particle volume, which is directly proportional to the blocking temperature, $T_B$, of the individual nanoparticles, $T_B \propto K_{ef}V$. Thus, the ratio of $T_B$s of the isolated nanoparticles quantifies the difference in energy barriers between the two types of particles. Therefore, we define an *anisotropy energy contrast*, AEC, as the ratio of the blocking temperatures of the two types of particles in the composite, $T_{B,HA}/T_{B,LA}$. Following this definition, the M1 series has a moderate AEC ≈ 1.6 (where the larger surface anisotropy contribution to $K_{ef}$ in the smaller particles explains the AEC < 2 (particles volume ratio)[44]), while the M2 series is characterized by a much larger AEC ≈ 8.2 (due to the large increase in anisotropy caused by the Co-doping of the maghemite particles).[45]

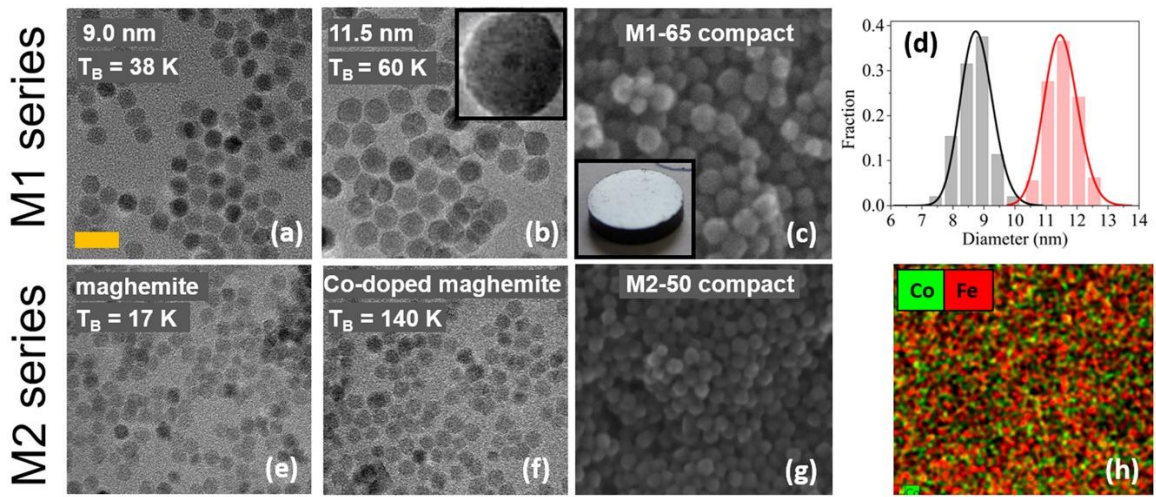

**Figure 1**. TEM images of the 9 nm (a) and 11.5 nm (b) particles used to prepare the M1-*x* samples. The inset in (b) shows a silica coated 11.5 nm maghemite nanoparticle. (c) Typical HRSEM image of a compact in the M1 series (*x* = 65), showing the nanoscale mixing of small and large particles. The inset shows one of the discs compacted under 1 GPa. (d) Size distributions of both types of particles. (e, f) TEM images of the 6.8 nm maghemite (e) and Co-doped maghemite (f) particles mixed in the M2 series. SEM image (g) and compositional mapping (h) of M2-50 proving the nanoscale mixing. All the images are scaled (orange scale bar = 20 nm). The TEM images in (a), (b), (e) and (f) were taken in the unmixed suspensions before removing the oleic acid coating. Note also that $T_B$ in the labels of panels (a), (b), (e) and (f) refers to the blocking temperature (defined as the peak temperature of the ZFC curve) of the isolated NPs (silica-coated).

The ZFC curves of the isolated nanoparticles, as well as those measured for all the M1 and M2 binary compacts, are shown in Figure 2. In both series there appears a single peak in the M(T) at $T_{MAX}$, suggesting that dipolar interparticle interactions (see SI, for a discussion on the nature of interparticle interactions)[46] are strong enough to provide collective behavior in the dense assemblies (*cf.* the superposition curves, corresponding to unmixed compacts, shown in Figure S1). In the M1 series the $T_{MAX}$ values are much higher than the individual blocking temperatures of both the LA and HA particles. This is consistent with the flat shape of the FC curve below the freezing temperature (as exemplified in the inset of Figure 2a for one of the composites), typical of strongly interacting particle systems. Note that although all samples freeze cooperatively at a single transition temperature, the broader effective size distribution in the central members of the series (as well as the demagnetizing field due to the disc shape in *all* samples)[47] smears out the transition.[48] On the other hand, for the M2 series, although the $T_{MAX}$ values are still higher than the blocking temperature of the hard NP (140 K), the increase is relatively small compared to the M1 series, hinting an important role for the particle anisotropy -and the particles anisotropy contrast- despite the single ZFC peak in the compacts. In fact, the ZFC curves also show a hump at low temperatures (T ≈ 30 K), except for the end members (see Figure 2b), a feature that can be more clearly observed in ac susceptibility measurements (see Figure S2). Notably, this effect is absent in the M1 series.

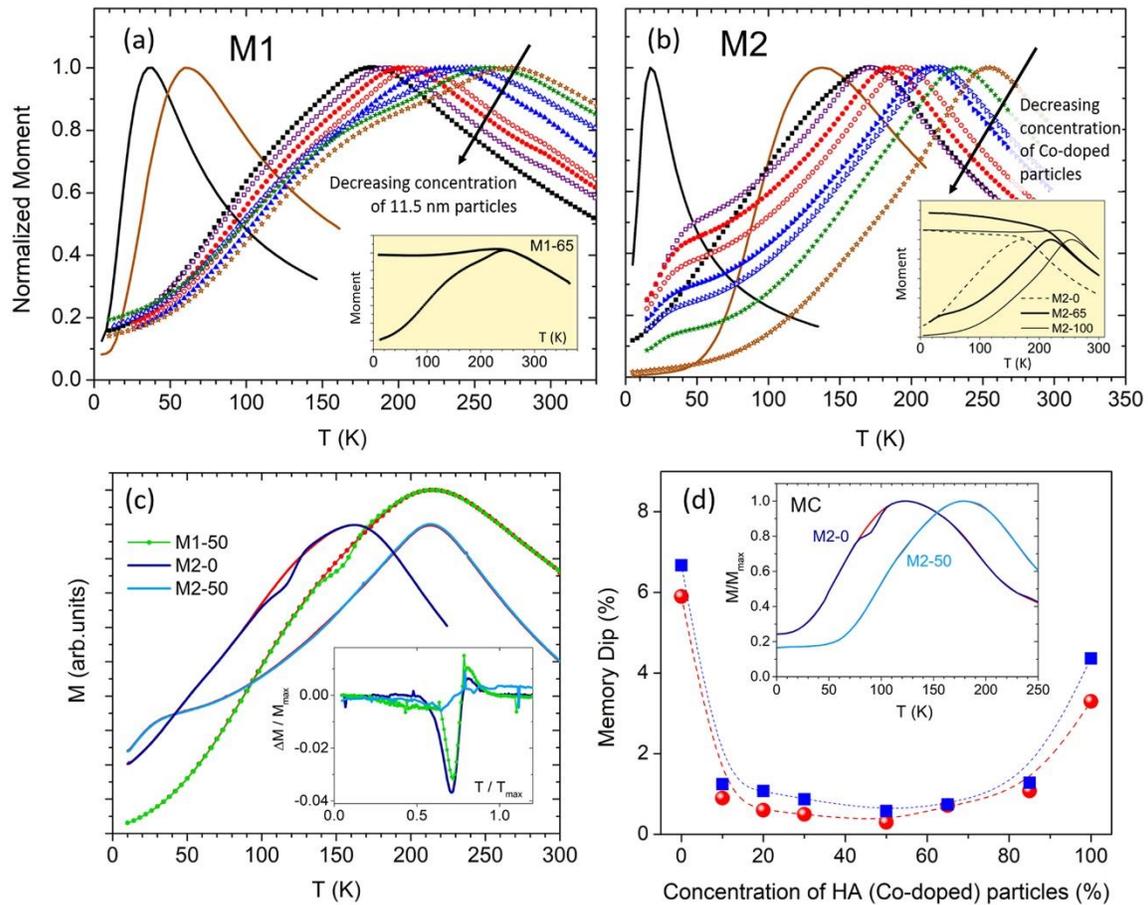

**Figure 2.** (a) and (b): zero-field-cooled normalized magnetization (M/M$_{MAX}$) curves measured in H = 5 Oe for the two series of compacts, as well as for the LA/HA particles isolated by silica spacers (black and brown solid lines, respectively). The insets show the field-cooled and zero-field-cooled curves for one of the mixtures of series M1 (all samples

in this series showed a very similar behavior) and for the end members as well as one of the mixtures in the M2 series. (c): ZFC memory experiments for M1-50, M2-0 and M2-50. The reference curves measured without a halt are plotted with red lines and the memory ZFC curves measured after a halt at $T_{halt} = 2 \cdot T_{MAX}/3$ are plotted with a green line for M1-50, a dark blue line for M2-0 and a light blue line for M2-50. The corresponding difference curves are plotted as a function of the reduced temperature $T/T_{MAX}$ in the inset. (d): Dependence on the concentration of HA particles of the ZFC memory dip in the M2 series, experimental (red circles) and Monte Carlo (blue squares) results. The inset shows two examples of simulated ZFC memory experiments (M2-0 and M2-50 systems).

To get a deeper understanding on the cooperative freezing of the samples, we have looked for a fingerprint of low-field collective behavior, namely the *ageing* and *rejuvenation* (leading to the ZFC *memory* effect) characteristic of the so-called *superspin glasses*, a state customarily observed at low temperature in strongly dipolar-interacting systems,[21,49,50] and originally examined in conventional spin glasses.[23,51] The memory effect manifests as a dip (with respect to a reference ZFC curve) at the halt temperature $T_{halt}$ in a ZFC "memory" curve measured in exactly the same conditions as the reference except for a halt at $T_{halt}$ during the zero-field cooling (see Methods). Remarkably, for the M2 series, we found that, although the end members of the series show a strong memory effect, this phenomenon is essentially suppressed in all the mixtures (Figure 2d), as exemplified in Figure 2c for the M2-50 sample. In contrast, all the samples in the M1 series show strong ZFC memory effects, as can be seen also in Figure 2c for the most unfavorable case, the M1-50 sample. Although still robust, the weaker memory effect in M2-100 compared to M2-0 is analogous to observations in conventional (atomic) spin glasses, where increasing spin anisotropies were shown to yield weaker memory effects.[52] On the other hand, the suppression of the memory effect with the introduction of even a small proportion of a (softer/harder) second phase is rather unexpected and highlights the crucial role of heterogeneity in the zero-field dynamics. This result was consequently explored by Monte Carlo simulations (using a three-spin model[53] for particles interacting exclusively through dipole-dipole interactions, see SI for details of the model), which reproduced precisely the experimental results (see Figure 2d, blue data points and example ZFC curves in the inset), both qualitatively (suppression of the effect upon mixing) and quantitatively (the values of the memory dip relative to the reference ZFC magnetization at $T_{halt}$).

The humps in the M(T) and the absence of memory effects in the M2 series are related to the difference in relaxation times of the pure and Co-doped particles (see Figure 3). Namely, the Co-doped particles are essentially blocked in the temperature range of the low temperature hump in the ZFC curves at the observation times of the experiments (~30 s for SQUID magnetization measurements). Thus, these particles act as weak static random fields, which are not able to participate in the dynamics of the system. In the M2-0 and M2-100, all the particles are equal, thus they all participate in the collective (equilibrium and non-equilibrium) dynamics. In the M2 mixtures, however, the blocked Co-doped particles affect the evolution towards an equilibrium phase and leave a fraction of the soft maghemite particles (not always the same) as quasi-superparamagnetic, yielding the low-temperature anomaly. Moreover, due to the much longer relaxation times of the Co-doped nanoparticles, they act as disturbances in the evolution towards equilibrium dynamics in the memory experiments. This implies that the system essentially remains in a non-equilibrium state in all time scales, thus the experiments probe non-equilibrium dynamics, which should have weaker memory effects. On the other hand, in the M1 series the relaxation times of the two types of particles are very similar (dashed lines in Figure 3), consequently, the dynamics of the different samples is more homogeneous leading to memory effects and the absence of a low temperature hump in M(T) for the whole series. Therefore, the presence of a single peak at $T_{MAX}$ in the ZFC curve measured in the mixtures, although resulting from strong enough interactions, does not necessarily imply single-phase dynamics below such peak temperature.

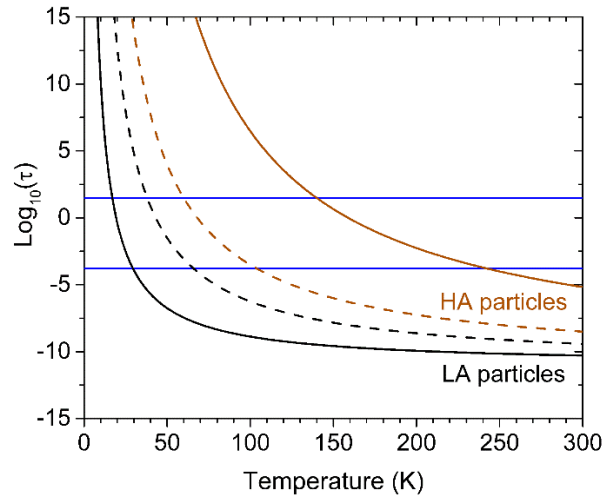

**Figure 3.** Evolution of the relaxation time (Arrhenius law) of the isolated LA/HA particles, considering $\tau_0 = 10^{-11}$ s and the blocking temperatures indicated in Figure 1, for the M1 (dashed lines) and M2 (solid lines) series. The horizontal lines mark typical observation times $\tau_{obs} \sim 30$ s (dc measurements) and $\sim 1/\omega$ (ac measurements, $\omega = 2\pi f$; $f = 10^3$ Hz).

This heterogeneity-driven suppression of dipolar collective behavior is similar to that found in systems with a *broad* particle size distribution (relative to the strength of the interactions), a typical form of "uncontrolled" heterogeneity.[34,54] Note that metallic NP systems allow for greater heterogeneity while preserving collective behavior due to the presence of strong non-dipolar interactions.[55,56] Thus, in the present study the control of the heterogeneity via the proportion of LA/HA has permitted to isolate the influence of this parameter from that of local anisotropy: e.g., M2-100 shows a smaller memory effect than M2-0 due to the much larger anisotropy of the NPs in the former sample, but the negligible memory in M2-50 (with a smaller average NP anisotropy than M2-100) must be then caused by the heterogeneity obtained by mixing the two types of particles.

Next, we examine the high-field behavior (hysteresis loops) of the two series. As can be seen in Figure 4, the end members of the M1 series have rather similar saturation magnetization, $M_S$ (although different magnetic moment, $\mu = M_S \cdot V$) and coercivity, $H_C$. On the other hand, the Co-doped NPs have a smaller $M_S$ than the pure maghemite NPs, but a much larger $H_C$. The smaller $M_S$ is probably due to increased disorder, presumably at the surface, as indicated by the observation of exchange bias (field-axis shift of the low temperature loop after cooling in a saturating field).[57,58] Interestingly, in contrast with the qualitatively similar trends of the ZFC curves for the two series, the hysteresis loops of the two types of mixtures are very different (Figure 4). While the M1 samples present ordinary (single-phase-like) loops, the M2 mixtures display double-loop responses typical of weakly coupled composites[13] or poorly-mixed (or phase-segregated) systems.[19,39] This was expected from the fact that the average dipole-dipole interaction strength in the M2-50 sample amounts to a field of ~140 Oe (after equating dipolar and Zeeman energies using the NP magnetic moments obtained as indicated in Figure S3), significantly lower than the difference in coercivity between the constituent particles (see Figure 4). In contrast, for M1-50 the corresponding average *dipolar field* is ~220 Oe, *i.e.*, larger than the coercivity difference between the LA and HA particles in series M1.

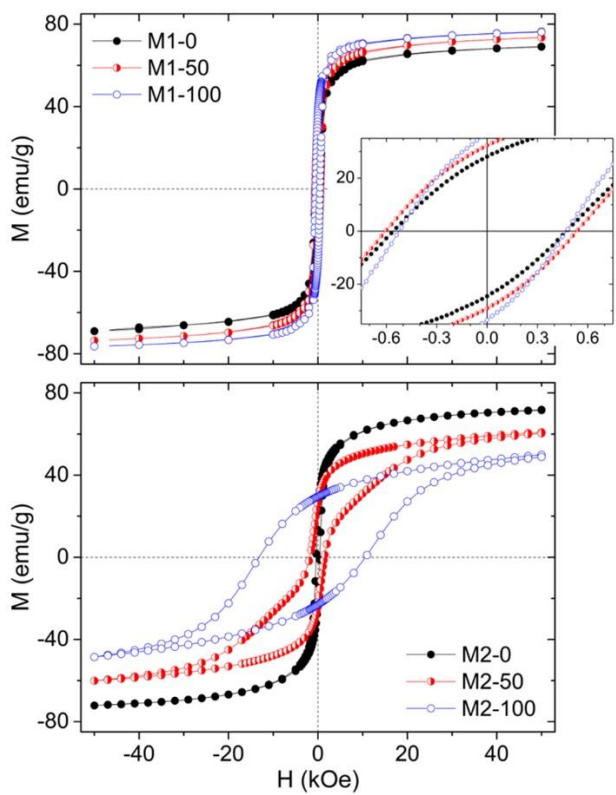

**Figure 4.** Hysteresis loops at 5 K after field cooling measured for the end and middle members of each sample series. The inset in the upper panel shows the low-field, hysteretic region.

Notably, although the overall shape of the measured loops is reasonably well-described as a superposition of the end member loops (see Figure 5), significant deviations appear at low fields for the central samples (see insets), yielding field-axis interception values larger than those extracted from the calculated superpositions. This result is, in fact, in agreement with the naïve picture of the harder particles providing some "pinning" against the switching of the softer particles, which essentially determine the overall $H_C$. Thus, one would intuitively argue that the difference in anisotropy between the hard and soft particles is high enough to prevent the full coupling of the two populations in the mixtures. However, the difference between the calculated and experimental coercivities indicate that the two types of NPs must influence each other through dipolar interactions. Thus, the soft/hard NP populations are strongly interacting (given the NPs proximity) but not fully-coupled due to the large anisotropy of the HA particles. Hence, our results highlight the ambiguity of the usual interpretation of the magnetic properties of a magnetic composite as "coupled" or "weakly coupled". To obtain the complete picture of the magnetic response, information on the field and temperature ranges should be discussed, as illustrated by the central samples in the M2 series, which show collective ("coupled") behavior at low fields and T ~ 200 K (*i.e.*, a single peak in the ZFC curve), but weak coupling at higher fields and T = 5 K (*i.e.*, separate, yet not independent, magnetization reversal of the two populations resulting in double-loop hysteresis).

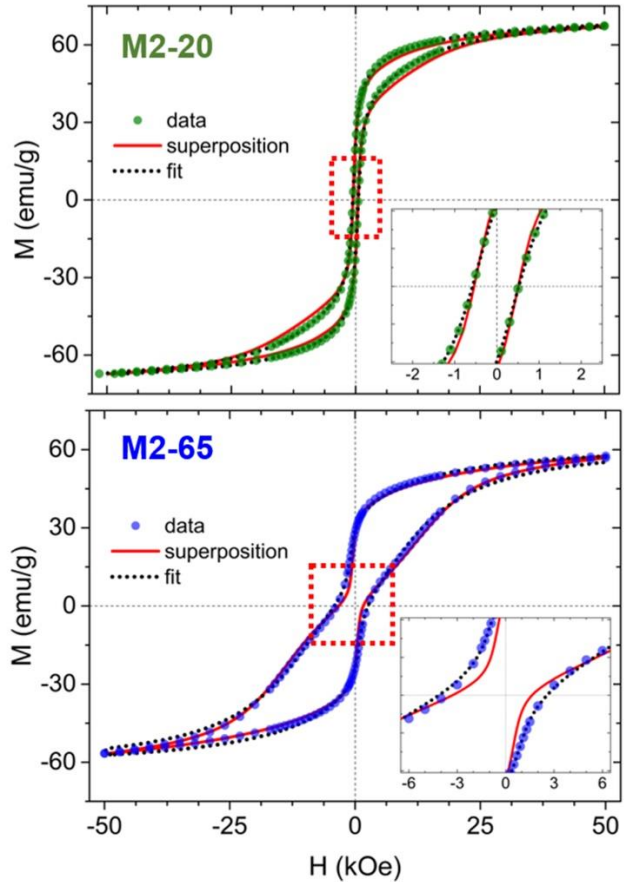

**Figure 5.** Hysteresis loops at 5 K after field cooling measured for two selected samples in the M2 series. The insets zoom the central region. The red line is the weighted superposition of the end members of the series (M2-0 and M2-100), whereas the black dotted line is the fit to a model with two interacting components (see text for details).

In the following we discuss in more detail the evolution of $T_{MAX}$ and $H_C$ on the proportion of HA particles, $x$, in the two series studied, as summarized in the two left columns in Figure 6. Firstly, we examine the data measured (or calculated) in the M1 series, *i.e.*, the left column in Figure 6. The freezing temperature ($T_{MAX}$) shows a clear linear trend, which can be understood by assuming that the individual blocking temperatures have no bearing on the collective freezing, *i.e.*, they are fully determined by the strong interparticle interactions, as first proposed by Mørup in the 90's[59] and suggested by the high $T_{MAX}/T_B$ ratio for the end members of the series. The data in the M1 series is well described by this model, where $T_{MAX}$ is proportional to the dipolar interaction strength between nearest neighbors $T_{dd}$,[38,59]

$$T_{MAX} \propto T_{dd} \propto \frac{\mu_{av}^2}{r^3} \propto M_S C \mu_{av} = M_S C [\mu_{LA} + (\mu_{HA} - \mu_{LA})x] \qquad (1)$$

where $\mu_{av}$ is the weighted average magnetic moment of the high- ($\mu_{HA}$) and low- ($\mu_{LA}$) anisotropy particles, and $C$ is the NP packing fraction/filling factor. Note that $C$ and $M_S$ are roughly constant across the series. A constant $C \approx 60\%$ results from the fact that all samples have been compacted under the same pressure, yielding quasi-random-close-packed configurations,[47] as shown in Figure 1c. Given the modest difference in diameter between the small (LA) and large (HA) particles, we do not expect a significant variation of the filling factor with the fraction of HA particles, $x$. The saturation magnetization of the large particles was measured to be only slightly larger than that of the small particles, as seen in Figure 4 upper panel.[44] Therefore, dipolar interactions depend on $x$ mainly through the volume-averaged particle moment, $\mu_{av}$, which varies linearly from the moment of the small particles to that of the large particles ($\mu_{LA}$ and $\mu_{HA}$, respectively). Thus, the experimental observation of a linear dependence of $T_{MAX}$ on $x$ confirms that the ZFC peak temperature is determined exclusively by the relatively strong dipolar interparticle interactions (*i.e.*, the relatively small NP anisotropies are irrelevant).

In these strongly coupled M1 mixtures, while $T_{MAX}$ is solely determined by interactions, $H_C$ is shown to be determined both by the intrinsic value in isolated NPs and by the intensity of the interparticle interactions. The $H_C$ values measured across the series deviate from the values extracted from the superposition loops, indicating an influence of the varying interaction strength. In fact, the observed non-monotonic dependence can be explained by a simple model which considers the variation across the series of both the average particle size and the interparticle interactions. The $H_C$ of a dense NP assembly results from two factors, the individual particle anisotropy barrier and the strength of interparticle interactions, which may be quantified by $T_{dd}$ (Equation 1). Regarding the first factor, it is well-known that the coercivity is proportional to the anisotropy barrier ($\propto KV$),[60] which, in turn (and given the similar K values previously measured for the particles in M1-0 and M1-100)[44] depends linearly on $x$. Thus, the *isolated-particle coercivity* of the average particle in the composite M1-$x$ can be written as

$$H_{C,ip}(x) = H_{C,LA} + (H_{C,HA} - H_{C,LA})x \tag{2}$$

where $H_{C,LA}$ and $H_{C,HA}$ are the coercive fields of the small and large particles, respectively. Regarding the effect of dipolar interactions, experiments dealing with differently concentrated dispersions of particles have established a decrease of $H_C$ with increasing particle concentration at temperatures well below the blocking temperature.[61] This is in agreement with the classical calculations by Néel[62] and Wohlfarth,[63] as well as with numerical simulations, which found a linear dependence between the two parameteres.[64] With $T_{dd} \propto C\mu_{av}$ for the strength of dipolar interactions in the M1 series, the present experiment is complementary to the cited previous experiments, as here the "concentration" $C$ (packing fraction) is constant and the average particle moment is finely tuned *via* the HA/LA proportion. Thus, the $H_C$ of the sample M1-$x$ can be written as

$$H_C(x) = H_{C,ip}(x)[1 - AT_{dd}(x)] \tag{3}$$

inserting $T_{dd}$ from Equation 1 yields

$$H_C(x) \propto \{H_{C,LA} + (H_{C,HA} - H_{C,LA})x\} * \{1 - B[\mu_{LA} + (\mu_{HA} - \mu_{LA})x]\} \quad (4)$$

where A and B are constants. This is an inverted parabola, as experimentally observed in series M1 (see Figure 6b1), which supports the mentioned approximations and hypotheses leading to Equation 4. Hence, intermediate compositions show a stronger effect of interactions (even if they increase monotonically in the series) and the coercivity of the uniform mixtures deviate more from the simple superposition of the two NP populations.

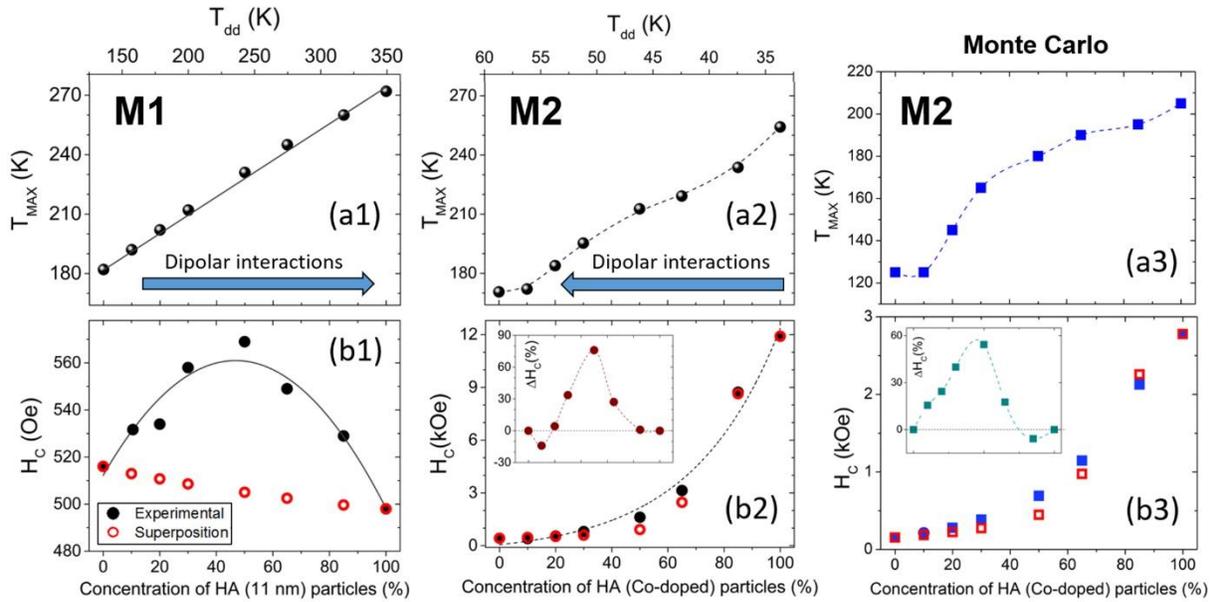

**Figure 6.** Dependence on the concentration of high anisotropy particles of the ZFC peak temperature, $T_{MAX}$ (a), and the coercivity, $H_C$ (b). The third column shows the results from Monte Carlo simulations for the M2 series. The empty red symbols in panels (b) correspond to values extracted from loops calculated as (weighted) superpositions of the loops measured (or simulated) for the end members of the series. The insets plot the relative difference between the values measured (or extracted from simulations) in the mixtures and in the superposition loops. The solid lines in the first column are fits to the models described in the text, while the dotted lines are guides to the eye. Note that the upper x-axes give the variation of the dipolar interaction strength, $T_{dd}$, as a result of the different magnetic moments of the LA and HA constituent particles. The arrows in (a1) and (a2) highlight the fact that while $T_{dd}$ increases with $x$ in the M1 series it decreases in the M2 series.

Next, we discuss the same magnetic parameters in the M2 series, plotted in the central column of Figure 6, and compare them to those in the M1 series. The dependence of $T_{MAX}$ on $x$ for the M2 series appears similar (increasing trend) to that of the M1 series. However, there is a crucial difference, highlighted by the upper axes in Figure 6a1 and a2, showing the variation of the average dipole-dipole interaction strength $T_{dd}$ with the mix proportion; namely, while $T_{dd}$ increases with $x$ in the M1 series, it decreases in the M2 series. This is because the magnetic moment of the Co-doped (HA) particles is smaller than that of the pure maghemite (LA) particles (see Figure S3 in the SI), thus $T_{dd}$ decreases (concomitantly with the increase in average local anisotropy) as the proportion of the lower-moment HA particles, $x$, becomes larger. Consequently, this rules out dipolar interactions as the origin of the increase in $T_{MAX}$ with $x$ and points out the significant influence of the local anisotropy on this characteristic temperature. Importantly, the data does not imply that dipolar interactions, although less intense than in series M1, are not also influencing the value of $T_{MAX}$. In fact, the existence of a single peak at this temperature can only be understood from the presence of strong enough interactions. Nevertheless, the average local anisotropy energy is larger and varies much faster across the series than the intensity of dipolar interactions: the ratio between the two energy terms ($KV/E_{dd} = 25T_B/T_{dd}$) ranging from roughly 7 (for M2-0) to about 10³ (in M2-100). Mørup's model, which was found to describe well the variation of $T_{MAX}$ in series M1, cannot account for $T_{MAX}(x)$ in the M2 series.

Regarding the coercive field, the experimental $H_C$ values also lie above the superposition values (panel b2 in Figure 6), as in the M1 series, but the enhancement due to uniform mixing is much larger (up to 80% for M2-50, see inset). This indicates that despite the double-loop behavior (signaling that the two populations are not fully coupled), the strong interaction between the particles significantly influence their magnetism. However, as discussed above, increasing concentrations of HA particles in this series provide *weaker* average dipolar interactions, therefore the model applied above for the M1 series is completely inadequate in the (relatively) weakly coupled M2 scenario, where the simpler intuitive idea of the harder particles "pinning" the switching of the softer ones, which in turn determine the overall coercivity, appears more suitable. Note that a minimum population of around 20% of HA particles is necessary for the soft particles to *feel* such pinning influence. The contrast between the two series in the influence of dipolar interactions on $H_C$ is thus remarkable.

The third column in Figure 6 shows results from Monte Carlo simulations of the M2 series. The concentration dependence of the parameters extracted from the simulated thermal and magnetic response (see SI for representative examples, Figures S5 and S6) are remarkably similar to the experimental observations, including the effect on the coercivity of mixing *versus* simple (unmixed) addition of the soft and hard nanoparticles [insets in panels (b2) and (b3)]. The relatively simpler, strongly coupled, M1 series was also simulated, with results again very similar to the experiment (see SI, Figure S4).

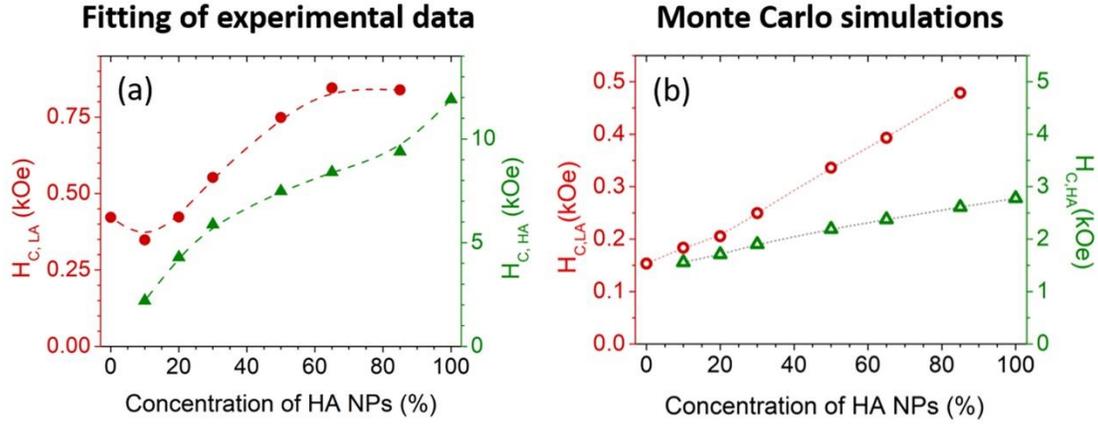

**Figure 7.** Coercive fields of the LA/HA NP populations as extracted from the fitting of experimental loops (a) and Monte Carlo-simulated loops (b) as a function of hard (Co-doped) particles concentration.

In order to explore the origin of the *x* dependence of the hysteresis loops in series M2 (summarized in panel b2 of Figure 6), we have attempted to extract the behavior of each population (LA and HA particles) by fitting the experimental loops to the following sum of *hard* and *soft* components using the empirical function proposed by Stearns and Cheng:[65]

$$M(H) = \frac{2}{\pi} M_{S,LA} \operatorname{atan}\left[ \left( \frac{H + H_{E,LA} \pm H_{C,LA}}{H_{C,LA}} \right) \tan\left( \frac{\pi S_{LA}}{2} \right) \right] + \\ \frac{2}{\pi} M_{S,HA} \operatorname{atan}\left[ \left( \frac{H + H_{E,HA} \pm H_{C,HA}}{H_{C,HA}} \right) \tan\left( \frac{\pi S_{HA}}{2} \right) \right] \quad (5)$$

where the ± symbol indicates the different sign used in the simultaneous fitting of the ascending and descending branches of the loops, and the squareness parameters (S) were obtained from the fitting of the end member loops (*i.e.*, it is assumed not to change significantly when mixing the two types of NPs). Note that the exchange bias of the loops is taken into account in the fit, $H_{E,LA}$ and $H_{E,HA}$. Two examples of the fitted curves are given in Figure 5. The results for $H_C$ as a function of the proportion of HA particles, *x*, plotted in the left panels of Figure 7, show a strong mutual influence of one type of particle on the other. Interestingly, the results are qualitatively reproduced by the Monte Carlo simulations (where it is straightforward to separate the individual contributions of the two particle populations to the total hysteresis loop). Thus, not only the hard particles *harden* the soft ones, as expected, by *delaying* their switching as commented above, but, conversely, the introduction of soft particles in a compact with majority of hard particles will *soften* them. This is the reason why the overall result of introducing hard particles in the binary compacts (increasing *x* in Figure 6b2) is a faster-than-linear enhancement of coercivity, as both components (LA and HA) are increasing their coercivity.

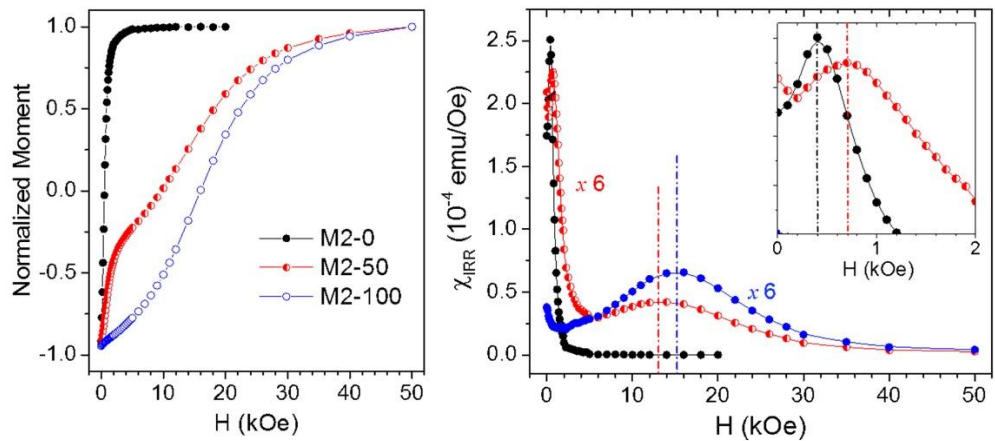

**Figure 8.** Switching field distributions (right panel) for the end and central members of the M2 series as obtained from the derivative of the DC demagnetization remanence ($M_{DCD}$) curves shown in the left panel. All curves were measured at a temperature of 5 K. The inset zooms in the low field region to show the deviation of the soft mode in the M2-50 mixture from the single peak observed in the M2-0 compact.

The relatively weak, but still significant, soft-hard particle coupling can be conveniently quantified by the analysis of the switching field distribution, defined as the field necessary to overcome the energy barrier during an irreversible reversal process (and therefore offering a measure of such anisotropy barrier distribution), which can be extracted from the dc demagnetization remanence ($M_{DCD}$) curve as $\chi_{irr} = dM_{DCD}/dH$.[39,66] Figure 8 shows the $M_{DCD}$ curves and corresponding switching field distributions for the end and central members of the M2 series. As expected from the overall aspect of its hysteresis loop, the central sample of the series, M2-50, shows two well-separated switching modes, soft (at low fields) and hard, which justifies the description of this composite as "weakly coupled". However, those modes are clearly shifted towards each other with respect to the switching behavior of the *pure* compacts M2-0 and M2-100, in agreement with the results from the Monte Carlo and fitting analyses above. It is this mutual influence that explains that the coercive field measured for sample M2-50 are considerably different from those of an unmixed system with the same components (see Figure 6, panels b2 and b3).

Therefore, despite the double-loop behavior of the hysteresis loop, Figure 7 and 8 clearly indicate that dipolar interactions between the HA and LA particles significantly affect their individual switching, suggesting the labels *modified single-particle response* or *weak coupling* to describe their behavior. The coupling, however, is high enough to render a single peak in the ZFC curve of the mixtures (see Figure 2b).

Altogether the results draw a complex scenario for the M2 mixtures, which present both seemingly "collective" ($T_{MAX}$) and "single-particle" ($H_C$) properties. In fact, the only "collective" property observed in these mixtures is the blocking temperature, although its increase with decreasing interactions and increasing local anisotropy already signals a dominant role for the latter factor. On the other hand, (i) the constricted hysteresis loops, (ii) the lack of ZFC *memory*, and (iii) the hump of the M(T) at low temperatures, they all correspond to a "two-phase" behavior driven by the large anisotropy contrast between the LA and HA particles.

The origin of the distinct difference between the M1 and M2 series must lie on the large anisotropy of the Co-doped maghemite nanoparticles. Based on the Random Anisotropy Model (RAM), in nanocrystalline materials the structural correlated volume, (*i.e.*, nanoparticle size) can actually be smaller than the volume of magnetically correlated material due to the coupling among particles/grains. Hence, within this framework, the magnetization reversal of such magnetic correlated volume is not ruled by the intrinsic properties of the individual particles, but by the averaged anisotropy and easy axis resulting from the sum of the individual contribution of each particle/grain randomly oriented within the "magnetic cluster".[67–70] For bulk nanocrystalline materials (*i.e.*, with exchange interacting grains), the region which influences the magnetization is given by the ferromagnetic correlation length, $L_{corr} = \sqrt{A_{ex}/K}$ (where K is the anisotropy constant and $A_{ex}$ is the exchange stiffness, which can be naïvely considered as the "force" trying to keep the spins parallel to each other). However, the RAM approach has been proposed to be independent of the source of the magnetic coupling and it should be appropriate for other types of interactions, like dipolar interactions.[68] Indeed, recent experimental observations have shown a good agreement with this model for ensemble of particles interacting purely by dipolar coupling.[71,72] However, for purely dipolar interacting systems, $A_{ex}$ is not the adequate parameter to define the tendency of neighboring magnetic moments to be correlated due to dipolar interactions. Thus, $A_{ex}$ can be qualitatively substituted by an "effective dipolar coupling stiffness", $A_{dip}$. Therefore, a (dipolar) correlation length can be estimated as $L_{corr} = \sqrt{A_{dip}/K}$. Thus, materials with a weak anisotropy, like $\gamma$-$Fe_2O_3$, should have a large $L_{corr}$, but Co-ferrite, with a large K, should have a small $L_{corr}$.

In fact, information on the intensity of the interparticle interactions can be extracted from the field dependence of $T_B$. Namely, using the well-known $T_B(H)$ for nanoparticles, $T_B = \frac{KV}{k_B \ln\left(\frac{\tau_m}{\tau_0}\right)}\left[1 - \frac{\mu_0 H}{\mu_0 H_K}\right]^{1.5}$, but taking into account the correlated volume (rather than the nanoparticle volume) and the effective, average, anisotropy of this correlated volume (rather than the intrinsic anisotropy of the nanoparticles); i.e., $V_N = \frac{\pi}{6}[D^3 + x(L_{corr}^3 - D^3)]$ (with D the particle diameter and $x$ the packing fraction), and $K_{eff} = \frac{K}{\sqrt{N}}$ (with N the number of particles contained in $V_N$), respectively.[70] See the SI for a more detailed derivation.[73,74]

By fitting the experimental $T_B(H)$ for the two pure cases of the M2 series, *i.e.*, M2-0 and M2-100 (see Figure S7) we obtain that, certainly, $L_{corr}$ for the $\gamma$-$Fe_2O_3$ is considerably larger than that obtained for the Co-doped $\gamma$-$Fe_2O_3$ particles (Table 1). In fact, $L_{corr}(\gamma$-$Fe_2O_3) = 36(1)$ nm encloses tens of nanoparticles (with D ≈ 6.8 nm) thus the magnetic properties of the pure maghemite particles in M2-0 are averaged over many particles. On the other hand, $L_{corr}$(Co-doped) = 11.4(5) corresponds to a correlated volume comprising barely 4 particles; consequently, the properties of the dense system made of Co-doped particles should be more individual-particle-like than those of M2-0, as observed experimentally.

Notably, the RAM approach described above has been developed for homogeneous mixtures of particles.[71,75] However, assuming that the approach holds even for binary mixtures, we can estimate the "average" $L_{corr}$ from the $T_B(H)$. Remarkably, the values obtained for the binary mixtures show that even a 10% of hard nanoparticles in the mixture is sufficient to reduce $L_{corr}$ to the M2-100 level.

**Table 1.** Correlation length ($L_{corr}$) for different samples of the M2 series obtained from the fit to a dipolar random-anisotropy model.

| Sample | $L_{corr}$ (nm) |
|---|---|
| M2-0 | 36(1) |
| M2-10 | 13.2(5) |
| M2-50 | 11.2(5) |
| M2-85 | 10.8(5) |
| M2-100 | 11.4(5) |

**CONCLUSIONS**

We have shown the convenience of dense hard/soft binary nanoparticle assemblies to discern the single-particle/collective nature of different properties in a given system. The influence of dipolar interactions, (average) local anisotropy, and sample heterogeneity on the low-field (blocking temperature, relaxation) and high-field (hysteresis loop) magnetic response has been illustrated in two series of strongly interacting, but strongly- (M1 series) and weakly-coupled (M2 series) composites resulting from different anisotropy ratios of the mixed hard/soft NP constituents.

The ZFC peak temperature ($T_{MAX}$) is a collective property in all samples studied. However, whereas in the M1 series $T_{MAX}$ is entirely determined by interparticle interactions, the M2 series presents a more complex scenario where the average local anisotropy provides a growing contribution with increasing concentration of the high anisotropy Co-doped NPs. Thus, the strong increase of the average anisotropy across this series (a factor of 8.2) overcomes the smaller reduction in dipolar energy (factor of ~1.6) to account for the observed increase in $T_{MAX}$. Nonetheless, it is the presence of strong interparticle interactions that enable, in the first place, the local averaging of anisotropy to yield a single stabilization temperature at $T_{MAX}$. However, in contrast with the M1 series, in the M2 mixtures the slow relaxation of the NPs moments below this temperature does not exhibit the ZFC memory effect characteristic of the (collective) superspin glass state, indicating a lack of homogeneous (single-phase) relaxation of the magnetization. More importantly, we have demonstrated the fundamentally different effect that dipolar interactions can have in nanoparticle composites depending on the anisotropy difference between the constituent NP populations. We have shown how binary random compacts with sufficiently high anisotropy contrast (*i.e.*, the M2 series) may be employed as a tool to test or, rather, define the *collective* character of a given magnetic property as that resulting in the collapse of the individual features caused by strong enough interactions. Crucially, such collective character must, in general, be ascribed to specific properties and not to the system as a whole.

**METHODS**

**Samples preparation.** Four types of highly uniform, roughly spherical NPs were synthesized using an optimized thermal decomposition route[58]: maghemite NPs with average diameters $d_{TEM}$ = 6.8, 9.0 and 11.5 nm, and cobalt-doped (Co:Fe = 0.24:1) maghemite particles 6.8 nm in diameter (see Figure 1). Iron pentacarbonyl (Fe(CO)$_5$) was thermally decomposed in the presence of oleic acid (surfactant) and dioctyl ether (solvent), and subsequently oxidized with trimethylamine N-oxide ((CH$_3$)$_3$NO) at high temperature. The nanoparticle size was controlled changing the amount of oleic acid in the reaction, e.g. for the 6.8, 9.0 and 11.5 nm particles, 1.7, 2.3 and 3.0 mol equivalents of oleic acid were used, respectively.[58] The Co-doped maghemite nanoparticles were prepared by simply replacing the corresponding fraction of Fe(CO)$_5$ by Co(CO)$_5$ to yield the above-mentioned Co:Fe ratio, which has been previously shown to produce a large increase in NP anisotropy while reducing only slightly the saturation magnetization.[45] The 9.0 and 11.5 nm NPs (corresponding to a volume ratio of 2), and the equally-sized, 6.8 nm, pure and Co-doped maghemite NPs, were mixed in different proportions while still in liquid solution to prepare the samples in series M1 and M2, respectively. The mixed solutions of nanoparticles were washed repeatedly in acetone to remove the oleic acid coating. Thermogravimetry analysis shows that an organic residue of only ≈ 5%w still remains bound to the NPs. The suspension was dried and the resulting powder compacted uniaxially under ≈ 1 GPa to yield dense discs with about 60% in filling factor, as estimated using a method based on the analysis of demagnetizing field effects recently developed by some of us.[47]

**Magnetic characterization.** The hysteresis loops were measured at 5 K after cooling in a 50 kOe field, which was also the maximum field used in the loops. The temperature dependence of the magnetization, M(T), (in H = 5 Oe) after field-cooling (FC) and zero-field cooling (ZFC) was also registered. In addition, *memory* ZFC curves were also measured. Namely, the cooling was halted during 4 hours at a given temperature, T$_{halt}$, below the ZFC peak temperature $T_{MAX}$ ($T_{halt} \sim 2 \cdot T_{MAX}/3$), then resumed to the lowest temperature (10 K). Subsequently, the M(T) curve is registered under exactly the same conditions as the reference ZFC curve. DCD (direct current demagnetization) remanence curves were measured by initially saturating the sample (in H = -50 kOe) and then measuring the moment after application and removal of progressively increasing reverse fields.[76] Finally, the temperature dependence of the ac susceptibility was recorded at 10 Hz using a field amplitude of 1 Oe. All the magnetic measurements were performed using a MPMS SQUID magnetometer from Quantum Design.

**Monte Carlo simulations.** Monte Carlo simulations were carried out using the mesoscopic three-spins model[53] (see Supporting Information for details).[77–79]

## ASSOCIATED CONTENT

Supporting Information Available: Extended magnetic experimental and theoretical data. Monte Carlo simulation model and random anisotropy model (RAM). This material is available free of charge via the Internet at http://pubs.acs.org.

## AUTHOR INFORMATION

**Corresponding Authors**

\* E-mail: joseangel.toro@uclm.es

\* E-mail: josep.nogues@icn2.cat

\* E-mail: elena.hsanchez@uclm.es

**ORCID ID**

Jose A. De Toro: 0000-0002-9075-1697

Josep Nogués: 0000-0003-4616-1371

Elena H. Sánchez: 0000-0001-5737-0035

**AUTHORS CONTRIBUTIONS**

J.A.T designed the experiment, J.N. expanded it to series M2; both coordinated the data analysis and discussion. S.S.L synthesized the nanoparticles and characterize them by TEM. E.H.S. and P.S.N. prepared the compacts, measured and analysed the M(T) and M(H) data. R.M, M.S.A and P.N are responsible for the memory measurements and analysis. M.V and K.N.T performed the Monte Carlo simulations. M.M. and P.C.R contributed to the magnetic characterization and performed the fittings of the two-phase hysteresis loops. G.S. analysed the mixtures homogeneity by SEM/EDS. G.M. and D.P. performed the RAM analysis. All authors contributed to the results discussion and revision of the article, which was written mainly by J.A.T., J.N and E.H.S.

**ACKNOWLEDGMENTS**

This work was supported by the Spanish Ministerio de Economía y Competitividad (grants MAT2015-65295-R and MAT2016-77391-R). JN also acknowledges funding from the Generalitat de Catalunya through the 2017-SGR-292 grant. ICN2 is funded by the CERCA Programme / Generalitat de Catalunya. The ICN2 is supported by the Severo Ochoa Centers of Excellence program, funded by the Spanish Research Agency (AEI, grant no. SEV-2017-0706). MSA, RM and PN acknowledge support from The Swedish Research Council (VR). The UCLM authors acknowledge technical help from Mario Rivera and Eduardo Prado.